\documentstyle[aps,prd,epsfig]{revtex}
\def\lsim{\raise0.3ex\hbox{$<$\kern-0.75em\raise-1.1ex\hbox{$\sim$}}}

\newcommand{\Tr}{\mathop{\rm Tr}}

\newcommand{\Dslash}{\hbox{D}\kern-0.6em\raise0.15ex\hbox{/}}

\preprint{          Preprint numbers: UUHEP 99/1 }
\begin{document}
\draft

\title{Scaling and Eigenmode Tests of the Improved Fat Clover Action}

\author{Mark Stephenson and Carleton DeTar}
\address{
Department of Physics, University of Utah, 
Salt Lake City, UT 84112, USA
}
\author{Thomas DeGrand and Anna Hasenfratz}
\address{
Department of Physics, University of Colorado, 
Boulder, CO 80309, USA
}
\date{\today}
\maketitle
\begin{abstract}
We test a recently proposed improved lattice-fermion action, the fat
link clover action, examining indicators of pathological
small-quark-mass lattice artifacts (``exceptional configurations'') on
quenched lattices of spacing 0.12 fm and studying scaling properties
of the light hadron spectrum for lattice spacing $a=0.09$ and 0.16 fm.
We show that the action apparently has fewer problems with
pathological lattice artifacts than the conventional nonperturbatively
improved clover action and its spectrum scales just as well.
\end{abstract}

\section{Introduction}

The goal of lattice fermion improvement schemes is to increase the
effectiveness of computer algorithms by significantly reducing  lattice
artifacts at economically coarse lattice spacings, particularly in the range
$a < 0.15$ fm.  The clover action with the clover coefficient
 chosen nonperturbatively
\cite{NPclover} (NPCA) has been found to give a
substantially better scaling of the hadron spectrum over the range
$[0.06,0.16]$ fm than the conventional Wilson fermion action (WA)
\cite{Gockeler,SCRI}.  However,  as
the lattice spacing is increased over this range, the NPCA becomes
increasingly sensitive to local fluctuations in the gauge
configurations that produce unwanted artifact singularities at small,
positive quark mass,
the so-called ``exceptional configurations.'' 
 To avoid them, one must keep the quark mass
artificially high.  At increasingly coarse lattice spacing, the lower
bound on ``safe'' quark masses rises, making an extrapolation to
physical quark masses increasingly problematical.  Difficulties with
such ``exceptional'' configurations are overcome in a variety of
approaches, including pole shifting \cite{BDEHT} and schemes which implement an
exact chiral symmetry on the lattice, the overlap
formalism \cite{overlap}, and the domain wall approach
\cite{domain_wall}.  Such methods are computationally expensive.
Here, we consider a new (approximate)
improvement scheme,  the ``fat clover''
action.

The ``fat clover'' action (FCA), proposed by DeGrand, Hasenfratz, and
Kov\'{a}cs \cite{Fat_Clover_Action}, couples the standard clover
action to a locally smoothed gauge field.  Smoothing of the gauge
fields is achieved through a series of APE blocking steps
\cite{APE_block}.  It is intended that the number of blocking steps
remains fixed as the continuum limit is approached.  Thus the
fermion-gauge coupling is modified at a scale that is a fixed multiple
of the cutoff, and the correct local action is recovered in the limit.
Smoothing has a number of beneficial effects: lattice artifacts are
suppressed, chiral properties are improved, and the renormalization of
a variety of lattice quantities, such as the local-vector-current
and axial-vector-current
renormalization constants $Z_V$ and $Z_A$, is small
 \cite{Fat_Clover_Action,DeGrandscaling}.

Here we examine two variants of fat link actions: one using the tree
level value for the strength of the clover term (TFCA) and one using
an optimized value (OFCA).  (See Table \ref{actions} for a guide to
our abbreviations.)

We first study the distribution of real eigenmodes of the lattice
Dirac operator in an ensemble of gauge configurations, with particular
attention to the ``exceptional'' eigenvalues at positive quark mass.
We show that the FCA has improved chiral properties, in the sense that
the spread of leading near zero modes is narrower than with the NPCA
at a lattice spacing of $0.12$ fm.  Thus the ``safe'' lower bound on
quark masses is lower with the FCA.  We then present a study of
fluctuations in the pion correlator as a qualitative test of the
suppression of exceptional configurations in NPCA and TFCA.

Finally we perform a small scaling test of the quenched hadron
spectrum for the TFCA and OFCA.  Scaling is tested as the lattice
constant is varied from $0.092$ fm to $0.164$ fm, choosing some fixed
value of the quark mass, such that either $m_\pi^2 = 2.5 \sigma$,
where $\sigma$ is the string tension, or $m_\pi/m_\rho \approx 0.7$
\cite{Fat_Clover_Action}.  Included for comparison are the
corresponding results for the standard Wilson action (WA),
nonperturbative clover action (NPCA), and the standard Wilson action
on a fat link background (FWA).  In this scaling study we pay careful
attention to the elimination of a variety of sources of systematic
error: we scale all quantities with physical dimensions and fix the
lattice dimensions in physical units.

In Sec.\ \ref{sec:action} we give details of the FCA and our
computational method.  In Sec.\ \ref{sec:results} we present results
of our lattice simulations.  We conclude with Sec.\
\ref{sec:conclusions}.

\section{Computational Method}
\label{sec:action}
%
\subsection{Fat Clover Action}
The fat clover action (FCA) is the usual clover action
\begin{equation}
  S_{SW} = \bar \psi M(\kappa)\psi =
S_W  - \kappa c_{SW} \sum_{\mu<\nu}\bar\psi(x) 
	\sigma_{\mu\nu} P_{\mu\nu}\psi(x) 
\end{equation}
where $S_W$ is the Wilson action, and
$P_{\mu\nu}$ is the standard ``clover'' expression for
the field strength tensor $F_{\mu\nu}$,
except that all link variables in $S_W$ and $P_{\mu\nu}$ are 
fat links. The fat link is constructed from the original (``thin'') links
 with a series of APE blocking steps \cite{APE_block}.  A
single step creates a new gauge configuration with each gauge link
replaced by a weighted sum of the link and its staples, followed by a
projection back to SU(3).  Explicitly, each link $U_{x,\mu}$ is
replaced by
\begin{equation}
V_{x,\mu} = {\cal P}\left[(1-c)U_{x,\mu} + \frac{c}{6}\sum_{\nu\ne\mu} \left(
U_{x,\nu}U_{x + \nu,\mu}U^{\dagger}_{x + \mu,\nu}
+U^{\dagger}_{x-\nu,\nu}U_{x - \nu,\mu}U_{x - \nu + \mu,\nu}
\right)\right].
\end{equation}
where ${\cal P}$ represents the projection, which in SU(3) chooses the
unique group element $U$ of SU(3) maximizing $\Tr(U V^\dagger)$.
(Replacement occurs after all smooth links for the lattice are
computed.)  This process is repeated $N = 10$ times in our study with
a coefficient $c = 0.45$.  Such a choice was found to give good
stability in instanton size and placement during smoothing in both
SU(2) and SU(3) \cite{rgmap_su2,rgmap_su3}.  We have not explored
other choices extensively, but suspect considerable latitude is
permitted in the choice of $c$ and $N$.  After smoothing, the mean
plaquette of the fat link is close to 1: For $6/g^2 = 5.7$, the mean
plaquette is ${\rm Tr}U_{plaq}/3= 0.985$ and for $6/g^2 = 6.0$ it is
$0.994$.

Fat link clover actions, like all clover actions, have no $O(a)$ nor
$O(g^2 a)$ corrections as long as the clover coefficient $c_{SW}$ is
taken to approach unity in the $g\rightarrow 0$ limit. In perturbation
theory\cite{Bernard} the effect of the fat link is to multiply the
usual thin link quark-gluon vertices by a form factor $(1-{c\over
3}a^2 \hat q^2)^N$ with $\hat q_\mu = (2/a) \sin(q_\mu a/2)$, where
$q$ is the gluon momentum.  This easily accounts for the observed
considerable reduction in additive quark mass renormalization and
finite renormalization constants for vector and axial currents close
to unity.

 The clover coefficient $c_{SW}$ is {\it a priori} unspecified.  We
expect, based on the near unit value of the mean plaquette, that an
optimal value for simulations would be close to the tree-level value
of unity, and that is what one would choose in a tadpole-based
improvement program\cite{LandM}.  We chose the value of $c_{SW}$ using
the approach of DeGrand, Hasenfratz and Kova\'cs, based on the
position of real eigenmodes of the Dirac operator
\cite{Fat_Clover_Action}.

  To locate the real eigenmodes, for each configuration we
calculated a noisy estimator of the expectation value
 $\langle \bar \psi \gamma_5 \psi \rangle$
\cite{Fat_Clover_Action,Bardeenresolve,Kuramashi}
\begin{equation}
  A(\kappa) = \overline{\eta}\gamma_5  M^{-1}(\kappa)\eta
\end{equation}
at a closely spaced series of real values $\kappa$, where $\eta$ is an
arbitrary random vector, held constant for the scan over $\kappa$.  A
real eigenvalue of $M(\kappa)$ appears as a pole in $A(\kappa)$,
provided the corresponding eigenvector has nonzero overlap with the
vector $\eta$.  The quantity $\bar \eta M^{-1}(\kappa)\eta$ would also
diverge at a real eigenvalue of $M(\kappa)$, but since the real
eigenmodes of $M(\kappa)$ are also close to being eigenmodes of
$\gamma_5$ with eigenvalue $\pm 1$, the factor $\gamma_5$ helps to
distinguish them.

In the continuum limit the only real eigenvalues of $M(\kappa)$ are
chiral zero modes, occurring at $\kappa_c = 1/8$, i.e. zero bare quark
mass.  If the lattice Dirac operator is not chiral, the real modes are
spread around $\kappa_c$ (defined as the value of $\kappa$ where
$m_\pi^2(\kappa)$ extrapolates to zero), and can also be shifted from
the chiral $\kappa_c=1/8$ value.  Such real eigenmodes are undesirable
lattice artifacts that prevent lattice simulations at small quark
mass. \cite{BDET}.  Configurations that produce them are called
``exceptional'', although the problem really lies with the choice of
fermion action and not with the gauge configuration itself.  Actions
with improved chiral properties have real eigenmodes that cluster more
closely around $\kappa_c$.  For such actions it should be possible to
study lower quark masses without encountering difficulties with
exceptional configurations, or to carry out a simulation on a coarser
lattice at the same quark mass.

For a sufficiently high degree of fattening one can optimize the 
clover coefficient by minimizing the spread
of the real eigenmodes on Monte Carlo-generated configurations.
In this work, however, we choose a simpler approach.  We generate 
a series of
artificial lattice instantons of varying size $a < r_0 < 3a$ on $8^4$
lattices and of size $2a < r_0 < 6a$ on $12^4$ lattices.  Instanton studies
at $6/g^2=5.7$ and $6/g^2=6.0$
predicted  instanton
sizes in this range \cite{markthesis}. For each such gauge
configuration, we examined the position of the resulting fermion real
eigenmode.  Results are shown in Tables \ref{small.tbl} and
\ref{large.tbl} and Fig.~\ref{fig:zero_vs_inst_size}. 
For large $r_0$ the near-zero modes are quite close
to zero quark mass.  As $r_0$ drops below the lattice cutoff, the
would-be zero mode moves toward negative quark mass.  The trajectory
of real eigenmodes is altered by adjusting the clover coefficient $c_{SW}$.
We find that with $c_{SW} = 1.2$ the variation in pole position is
minimized for instanton sizes in the range $r_0 > a$.  With $c_{SW} =
1.1$ the variation in pole position is minimized for $r_0 > 2a$.
Since our scaling test considers lattices over a range of spacings
varying by a factor of 2, we choose $c_{SW} = 1.2$ for our coarsest
lattice spacing and $c_{SW} = 1.1$ for our finest lattice spacing to
assure scaling consistency.  We call the action with tuned clover
coefficient the ``optimized fat clover action'' (OFCA).  To test
sensitivity to this choice, we also present results with the
tree-level choice $c_{SW} = 1$ (TFCA).

\subsection{Computational Parameters}
There are two parts to this study: an analysis of the distribution of
low energy real eigenvalues and an analysis of the scaling of the
spectrum.  For studies of the distribution of near-zero eigenvalues at
$6/g^2 = 5.85$ we analyzed 100 configurations of size $10^4$, and for
the companion spectrum study to relate the pion mass to the $\kappa$
value, 20 configurations of size $12^3 \times 48$.  For the spectrum
scaling studies we have worked with two ensembles of quenched gauge
configurations generated with the conventional one-plaquette Wilson
action: 120 configurations of size $8^3\times 24$ at $6/g^2 = 5.7$ and
100 of size $16^3 \times 48$ at $6.0$, corresponding to lattice
spacing $a = 0.164$ and $0.092$ fm, respectively, based on recent
measurements of the string tension for this action in lattice units
\cite{SCRIstring} and the choice $\sqrt{\sigma} = 468$ MeV
\cite{Bali}.

The lattice dimensions for the scaling study were chosen to keep an
approximately constant physical volume, so as to avoid inconsistent
finite size effects. To allow tuning of the quark mass, either by
fixing the pion mass in terms of the string tension or in terms of the
rho mass, we calculate the spectrum for (typically) three neighboring
$\kappa$ values selected so that the desired dimensionless ratios can
be reached by interpolation.  Quark propagators were generated from a
fixed Gaussian (shell-model) source with standard deviation $2a$ for
the coarsest lattice and $4a$ for the finest.  Our scaling tests were
computed at three values of the clover coefficient $c_{SW}$ for each
gauge coupling: $c_{SW} = 1.1$ or 1.2, as noted above, for the OFCA,
$c_{SW} = 1$ for the tree-level fat clover action (TFCA), and $c_{SW}
= 0$ for the fat Wilson action (FWA) to test the relative merits of
smoothing and reducing the O(a) errors in the action.
%
\section{Results}
\label{sec:results}
\subsection{Fermion eigenmodes}
For each of the actions in our study we determined the distribution of
the real eigenmodes on a set of 100 $10^4$ gauge configurations,
generated with the conventional single plaquette action at $6/g^2 =
5.85$.  We determined the probability distribution $P(m_\pi^2)$ of the
leading pole (i.\ e. the eigenvalue corresponding to the largest quark
mass) for the various actions in our study.

Note that this statistic is different from the eigenvalue histograms
of Ref.~\cite{Fat_Clover_Action}, where {\em all} the low energy
eigenmodes were included to study the spread of the physical modes.
Since here we consider only the pole corresponding to the largest mass
on each configuration, these plots are indicative of the exceptional
configurations.

To compare leading-pole distributions from the various actions, we
converted the $\kappa$ values to $m_\pi^2$ values, by measuring the
hadron spectrum with the same action on a set of 20 $12^3 \times 48$
quenched configurations.  Results are shown in
Fig.~\ref{fig:prob_leading}.  Pion masses used for constructing the
linear scale conversion, $m_\pi^2 = a/\kappa + b$ are given in Tables
\ref{kappa_scale_a} and \ref{kappa_scale_b}.  The bin widths in this
figure are variable, since the poles were located by scanning at the
same constant increment in $\kappa$ for all actions.  We took this
interval as the resolution of the pole location.  The corresponding
interval in $m_\pi^2$, however, varied from action to action.  The bin
heights are scaled so that the probability distributions all have unit
total area.  The TFCA and OFCA actions clearly produce distributions
that are more sharply clustered around $m_\pi^2 = 0$.  The peak for
the OFCA appears at a nonzero bin in $m_\pi^2$, but we estimate a
combined systematic and statistical error (one sigma) of one bin width
arising from the conversion from $\kappa$ to $m_\pi^2$ near zero pion
mass.  

To put these results in another perspective, we have also measured the
ratio $m_\pi/m_\rho$ for these configurations.  For our sample of
gauge configurations at $\beta=5.85$ the NPCA encounters its first pole at
$m_\pi/m_\rho = 0.56(4)$, the TFCA, at about 0.37(6) and the OFCA at
about 0.42(5).

\subsection{Pion Correlators}

While a pole in the quark propagator at $\kappa < \kappa_c$ is the
most precise indicator of an exceptional configuration, fluctuations
in a hadron correlator, e.g.\ the pion correlator, give a qualitative
indication.  Poles in the quark propagator typically have residues
with concentrated support in Euclidean space-time.  For
instanton-induced poles, localization of the residue comes from
localization of the zero eigenmode.  Since the pion propagator is the
gauge-invariant square of the quark propagator, a nearby exceptional
pole typically contributes a strong localized fluctuation in the pion
correlator$-$in some cases inducing a ``W'' shape in a semilog plot.

Starting from a common sample of 80 quenched single-plaquette $8^3
\times 24$ gauge configurations at $6/g^2 = 5.7$, we compute the pion
correlator for the NPCA and TFCA for a range of quark masses.  We then
consider two measures of fluctuations:  (1) the noise to signal
ratio of the correlator at a fixed time and 
 (2) the number of outliers,
based on a correlated chi square measure.

The simplest measure of fluctuations is the noise to signal ratio in a
correlator, namely the ratio of the standard deviation of a correlator
$\sigma(t)$ to the value of the correlator, $\bar c(t)$, as a function
of distance and/or quark (or pion) mass.  The pseudoscalar correlator
is the most useful one to look at, since simple theoretical arguments
suggest that $\sigma(t)/\bar c(t)$ should be roughly independent of
the quark mass.  In Fig.~\ref{fig:noise_to_signal} this ratio at $t =
10$, namely $\sigma(10)/\bar c(10)$, is plotted over a range of
$m_\pi/m_\rho$ for the NPCA and TFCA.  It is clear that for $0.5 <
m_\pi/m_\rho < 0.7$, fluctuations are dramatically reduced with the
TFCA.

We denote the correlator on the $i$th configuration by $c_i(t)$ and
its mean over the sample of configurations by $\bar c(t)$. For the
outlier test we start by constructing the usual covariance matrix
$v_{t,t^\prime}$, based on the observed fluctuations, and its inverse
$w_{t,t^\prime}$.  The correlated chi square measure for configuration
$i$ is then
\begin{equation}
  \chi^2_i = \sum_{t,t^\prime}[c_i(t) - \bar c(t)]
     [c_i(t^\prime) - \bar c(t^\prime)] w_{t,t^\prime}
\end{equation}
for $N_t$ degrees of freedom.  The corresponding confidence level is
then used to determine the strength of deviation from the mean.  For
the sample of 80 configurations we treated the two time intervals $[0,
N_t/2]$ and $[N_t/2+1,N_t-1]$ separately.  A configuration was deemed
exceptional, if the confidence level determined on either interval was
less than $10^{-7}$, a somewhat arbitrary value that was chosen to
correspond to strongly discernible deviations, many of them with the
``W'' shape, characteristic of an exceptional configuration.  Note
that a Gaussian normal fluctuation at this level in a sample of 80
would be expected only once in about $10^5$ trials. The process of
identifying outliers was carried out iteratively, in each pass
removing the outliers from the sample as they were identified, until
none remained.  Results are summarized in Tables \ref{outliers_NPCA}
and \ref{outliers_TFCA}.  Based on this measure, if we were to insist
on no more than one exceptional configuration in a sample of this
size at $\beta=5.7$, the NPCA would be restricted approximately to
 $m_\pi/m_\rho >
0.8$ and the TFCA to $m_\pi/m_\rho > 0.6$.

\subsection{Spectrum}
Correlators for the zero momentum pion, rho, nucleon, and $\Delta$
were fit to single exponential forms, minimizing the correlated
$\chi^2$.  Care is needed to prevent biases arising from the choice of
the fitting range, particularly from the choice of minimum time
$t_{\rm min}$.  We used two methods to test for bias: In fitting
correlators for a set of closely spaced $\kappa$ values we selected
(1) the smallest $t_{\rm min}$ giving a minimum CL for all $\kappa$'s
greater than 0.05 and an average CL greater than 0.1; and (2) the
$t_{\rm min}$ for which the product of CL and the number of degrees of
freedom (df) is maximum, a rather ad hoc rule of thumb \cite{HEMCGC}.
As a rule both methods gave the same $t_{min}$.  Where a different
value was obtained, we determined that the variation in mass value was
within the statistical errors of the fits.  

We found that $t_{\rm min}$ for the $8^3 \times 24$ lattice was
generally half the value on the $16^3 \times 48$ lattice.  In the few
cases in which it was not (the rho meson for OFCA and TFCA), we
verified that, had we enforced this further condition, the central
mass value would have shifted by less than 1\%, an amount smaller than
the error in the observed scaling violation.  Our fitting range is
then approximately constant in physical units, and our results are
therefore free of bias from this source.

We also checked the single exponential fits against two-exponential
fits and verified that the results were stable within statistical
errors.  Results of the fits are shown in Table \ref{summary.tbl}.

We consider two alternatives for fixing the quark mass: (1) fixing
$m^2_\pi/\sigma = 2.5$ and (2) fixing $m_\pi/m_\rho = 0.7$.  These
values were chosen to correspond to each other, approximately.  Since
variations in the strength of the clover term changes the $N-\Delta$
and $\pi-\rho$ mass splittings, so can change the pi to rho mass ratio
at fixed physical quark mass, the former method is preferable. 
We present the 
second, more popular, method to allow comparison with other work.

Table \ref{scalingpi.tbl} and Figs.~\ref{fig:scalingpimasses} and
\ref{fig:scalingpiratios} show the masses of particles and their
ratios at $m_{\pi}^2 = 2.5 \sigma$ and the scaling violations from
$6/g^2=5.7$ to $6/g^2=6.0$.  For comparison, conventional Wilson data
(WA) from recent calculations are shown.  The WA $6/g^2 = 5.7$ values
are interpolated from raw data given in Ref. \cite{wilson3}.  The WA
$6/g^2 = 6.0$ values are interpolated from raw data given in Refs.
\cite{wilson2,wilson1}.  The value $m_{\pi}^2 = 2.5 \sigma$ is
slightly outside the range of values given in Ref. \cite{wilson2}, so
to avoid extrapolation, those data were supplemented by data given in
Ref. \cite{wilson1}; however, extrapolation from the data of
Ref. \cite{wilson2} alone gives the same results.
The $6/g^2 = 5.7$ NPCA values were interpolated from unpublished
values provided by Heller \cite{Heller_unpub}.  The $6/g^2 = 6.0$ NPCA
values are from Ref.~\cite{Gockeler}.

The mass of the rho meson is seen to be a sensitive indicator of
scaling violations.  The scaling violation in the rho mass is reduced
from approximately 7\% for the WA to less than 2\% for the OFCA and
TFCA. 
 Fattening the WA does not improve scaling with any
significance.  Thus smoothing of the gauge fields alone does not
improve scaling. This is because the Wilson action, with
 either thin or fat links, has $O(a)$ lattice artifacts which are removed by
the addition of the clover term. (A similar behavior for a fat link action
without a clover term, in that case, a hypercubic action, was also
seen in Ref. \cite{DEGRAND}.)

Scaling violations of the nucleon mass are statistically consistent
for all the actions considered.  Scaling violations of the delta mass
are improved from roughly 8\% for the FWA to less than 4\% for the
TFCA and OFCA.  This also is found for the alternative fits of the
delta masses, suggesting that the scaling improvement seen is real,
despite potentially large systematic errors in the fits of the delta
masses.

We extrapolated the mass values to zero lattice spacing, forcing a
common extrapolated mass for all actions.  Our extrapolation is linear
in $a$ for the WA and FWA and linear in $a^2$ for the OFCA, TFCA and
NPCA\@.  Results are plotted in Figs.~\ref{fig:scalingpimasses} and
\ref{fig:scalingpiratios}.  It is clear that none of the actions
completely remove scaling violations in the nucleon or delta mass, but
that all of the clover actions show  smaller violations than the Wilson
actions.

For the second approach we adjust quark masses so as to fix the ratio
$m_\pi / m_\rho = 0.7$.  Table \ref{scalingpirho.tbl} shows the masses
of particles and their ratios, and the scaling violations from
$6/g^2=5.7$ to $6/g^2=6.0$.  For comparison, conventional Wilson data
from recent calculations are shown.  The Wilson $6/g^2 = 5.7$ values
are interpolated from raw data given in Ref. \cite{wilson3}.  The
Wilson $6/g^2 = 6.0$ values are interpolated from raw data given in
Ref.  \cite{wilson2}.  Also shown are data for the TFCA and NPCA.  For
the NPCA the $6/g^2 = 5.7$ values are given in Ref.  \cite{SCRI},
already interpolated to $m_\pi / m_\rho = 0.7$.  The values at $6/g^2
= 6.0$ are from Ref. \cite{Gockeler}.

The mass of the rho again is seen to be a sensitive indicator of
scaling violations.  For the pair of lattice spacings used, the
scaling violation of the rho mass is reduced from approximately 13\%
for the conventional WA to less than 2\% for the OFCA or the TFCA.
There is no significant scaling improvement of the rho mass for the
FWA compared with the WA.  The scaling violations of the rho are
compounded with those required by the fixing of the pion to rho mass
ratio to the same constant for all cases.  The quark masses are forced
to values at $6/g^2 = 5.7$ and $6/g^2 = 6.0$ such that the scaling
violation of the pion mass equals that of the rho mass.  This roughly
doubles the total rho scaling violation at fixed $m_\pi / m_\rho$,
compared with that for fixed $m_{\pi}$, for cases where scaling
violations in the pi-rho mass splittings are large, as will be shown
to be the case for the Wilson actions.

Scaling violations of the nucleon mass also are compounded with those
required by the fixing of the pion to rho mass ratio to the same
constant for all cases.  The value of the nucleon mass is lowered
similarly to the pion mass by the clover (magnetic moment) term, so
forcing the scaling violation of the pion mass to equal that of the
rho mass also forces the nucleon mass to acquire a similar scaling
violation.  At $m_\pi / m_\rho = 0.7$, scaling violations of the
nucleon mass are reduced from approximately 10\% for the WA and FWA to
approximately 5\% for the TFCA to approximately 2\% or lower for the
OFCA.

Scaling violations of the delta mass are improved from roughly 14\%
for the FWA to less than 5\% for the OFCA.

Mass splittings are given in Table \ref{splittings.tbl} at $m_{\pi}^2
= 2.5 \sigma$.  Phenomenological values of the differences of the
squared masses of vector and pseudoscalar particles are almost equal
for different quark flavors, so the unrealistically high quark mass
used in the lattice calculations should not matter much.  Similarly,
the difference of the masses of spin $\frac{1}{2}$ and spin
$\frac{3}{2}$ baryons are comparable for different flavors.  Scaling
violations of the mass splittings are reduced substantially for the
clover actions compared with the Wilson actions.  Also there is rough
agreement with experimental values for the clover actions.  The
experimental values are $(m_{\rho}^2-m_{\pi}^2) = 5.7 \times 10^5$ MeV$^2$ and
$m_\Delta - m_N = 294$ MeV.  Using $\sqrt{\sigma} = 468$ MeV, we have
$(m_{\rho}^2-m_{\pi}^2)/\sigma = 2.6$ and $(m_N-m_\Delta) /
\sqrt{\sigma} = 0.63$.

In summary, the proposed fat clover actions with either tree-level
clover coefficient (TFCA) or optimized clover coefficient (OFCA), have
greatly improved scaling properties compared with the Wilson action.
The scaling improvement with either of these actions is comparable to
that of the NPCA.  Scaling tests of the FWA show that smoothing by
itself does not improve scaling. The combination of the clover term
with a fat link allows one to reach smaller values of the pseudoscalar
mass than is possible with the NPCA, with an apparently equivalent
level of scaling violations.

\section{Conclusions}
\label{sec:conclusions}

Using a common set of gauge configurations, we have carried out a
systematic study of the distribution of leading near-zero eigenvalues
and the scaling of the light hadron spectrum and for a variety of
fermion actions on quenched lattices with lattice spacing in the range
$[0.09,0.16]$ fm.  Actions included in this study are the conventional
Wilson action (WA), Wilson action on fat gauge links (FWA), clover
action with a non-perturbatively tuned clover coefficient (NPCA),
tree-level ``fat'' clover action (TFCA) and optimized fat clover
action (OFCA). All of the clover actions show better scaling behavior
than any of the Wilson actions--an entirely expected result, since
adding the clover term with the correct coefficient converts an $O(a)$
action into an $O(a^2)$ one.  Based on an analysis of pole positions,
we have found that the fat clover actions OFCA and TFCA exhibit chiral
properties superior to the NPCA, WA, and FWA.  A further analysis of
fluctuations in the pion correlator shows that the TFCA is far less
noisy than the NPCA, a further circumstantial indication of a
suppression of exceptional configurations.

\acknowledgements

We thank Urs Heller for providing us with unpublished results for the
nonperturbative clover action and for a careful reading of the
manuscript.  This work is supported in part by the U.S.~National
Science Foundation under grants PHY 96--01227 and PHY 99--70701
and by the U.~S. Department of Energy under contract
DE-FG03-95ER40894.
Computations were carried out with grants of computer time for the T3E
at the San Diego Supercomputer Center and for the IBM SP at the
University of Utah Center for High Performance Computing.

%

%

\begin{table}
\caption{Guide to abbreviations for the fermion actions in this study.
\label{actions}}
\begin{tabular}{ll} \hline
FCA  & Clover action on fat gauge links (either optimized or tree-level). \\
OFCA & Clover action on fat gauge links with optimized clover coefficient. \\
TFCA & Clover action on fat gauge links with tree-level clover coefficient. \\
NPCA & Conventional clover action with non-perturbatively tuned clover coefficient.\\
FWA  & Wilson action on fat gauge links. \\
WA   & Conventional Wilson action ($r = 1$). \\
\hline
\end{tabular}
\end{table}

\begin{table}
\caption{Pole location $m_p a$ for small instantons\label{small.tbl}}
\begin{tabular}{ddddddd} \hline
\multicolumn{1}{c}{} &
\multicolumn{6}{c}{$c_{SW}$}\\
$\rho / a $ & 1.0 & 1.1 & 1.2 & 1.3 & 1.5 & 1.8 \\
 \hline
1.00 & $-$0.141 &$-$0.073 & $-$0.003 & 0.071 & $-$ & $-$ \\
1.25 & $-$0.083 &$-$0.036 & 0.012 & 0.065 & 0.188 & $-$ \\
1.50 & $-$0.050 &$-$0.013 & 0.021 & 0.061 & 0.146 & $-$ \\
1.75 & $-$0.030 &$-$0.001 & 0.025 & 0.056 & 0.121 & $-$ \\
2.00 & $-$0.019 & 0.004 & 0.026 & 0.050 & 0.102 & 0.194 \\
2.25 & $-$0.012 & 0.006 & 0.025 & 0.045 & 0.088 & 0.159 \\
2.50 & $-$0.009 & 0.007 & 0.023 & 0.041 & 0.078 & 0.138 \\
2.75 & $-$0.007 & 0.007 & 0.021 & 0.037 & 0.070 & 0.122 \\
3.00 & $-$0.006 & 0.006 & 0.020 & 0.034 & 0.064 & 0.111 \\
 \hline
\end{tabular}
\end{table}

\begin{table}
\caption{Pole location $m_p a$ for larger instantons\label{large.tbl}}
\begin{tabular}{dddd} 
\hline
\multicolumn{1}{c}{} &
\multicolumn{3}{c}{$c_{SW}$}\\
$\rho / a $ & 1.0 & 1.1 & 1.2 \\
 \hline
2.00 & $-$0.019 &    0.003 &    0.024 \\
2.50 & $-$0.008 &    0.005 &    0.020 \\
3.00 & $-$0.004 &    0.005 &    0.015 \\
3.50 & $-$0.002 &    0.005 &    0.012 \\
4.00 & $-$0.002 &    0.004 &    0.010 \\
4.50 & $-$0.002 &    0.003 &    0.009 \\
5.00 & $-$0.002 &    0.003 &    0.007 \\
5.50 & $-$0.002 &    0.002 &    0.007 \\
6.00 & $-$0.003 &    0.002 &    0.006 \\
 \hline
\end{tabular}
\end{table}
\begin{table}
\caption{Pion masses (smeared-local channel) in lattice units vs
$\kappa$ for a variety of actions on quenched $12^3 \times 48$
lattices at $6/g^2 = 5.85$.  \label{kappa_scale_a}}
\begin{tabular}{dd|dd|dd} \hline
\multicolumn{2}{c}{OFCA} & \multicolumn{2}{c}{TFCA} & 
\multicolumn{2}{c}{NPCA} \\
\hline
0.1200 & 0.621(6)  & 0.1220 & 0.539(4)  & 0.1300 & 0.733(7)  \\  
0.1210 & 0.547(6)  & 0.1230 & 0.463(4)  & 0.1310 & 0.645(7)  \\  
0.1220 & 0.465(7)  & 0.1240 & 0.378(5)  & 0.1320 & 0.547(8)  \\  
0.1230 & 0.370(11) & 0.1245 & 0.319(5)  & 0.1330 & 0.432(9)  \\  
0.1235 & 0.316(17) & 0.1250 & 0.260(12) & 0.1335 & 0.364(10) \\
\hline
\end{tabular}
\end{table}
\begin{table}
\caption{Continuation of table \protect\ref{kappa_scale_a}\label{kappa_scale_b}}
\begin{tabular}{dd|dd} \hline
\multicolumn{2}{c}{FWA} & \multicolumn{2}{c}{WA} \\
\hline
0.1265 & 0.515(6)  & 0.1560 & 0.511(9)  \\  
0.1275 & 0.455(6)  & 0.1570 & 0.457(10)  \\  
0.1285 & 0.389(7)  & 0.1580 & 0.400(12) \\  
0.1295 & 0.309(10) & 0.1590 & 0.336(16) \\  
0.1300 & 0.261(13) & 0.1595 & 0.298(18) \\
\hline
\end{tabular}
\end{table}
\begin{table}
\caption{Outliers at CL  $<10^{-7}$ for the NPCA in a sample of 80.
\label{outliers_NPCA}}
\begin{tabular}{ddd} \hline
$\kappa$ & $m_\pi/m_\rho$ & $N$ \\
\hline
0.125  & 0.84(1) & 0  \\  
0.127  & 0.78(1) & 4  \\  
0.129  & 0.68(1) & 5  \\  
0.1295 & 0.64(1) & 7  \\  
0.130  & 0.48(2) & 12 \\
0.1302 & 0.50(2) & 15 \\
0.1303 & 0.51(3) & 14 \\
\hline
\end{tabular}
\end{table}
\begin{table}
\caption{Outliers at CL $<10^{-7}$ for the TFCA in a sample of 80.
\label{outliers_TFCA}}
\begin{tabular}{ddd} \hline
$\kappa$ & $m_\pi/m_\rho$ & $N$ \\
\hline
0.121 & 0.80(1) & 0 \\
0.122 & 0.75(1) & 0 \\
0.123 & 0.70(2) & 0 \\
0.124 & 0.62(2) & 0 \\
0.125 & 0.50(2) & 2 \\
0.126 & 0.42(4) & 5 \\
\hline
\end{tabular}
\end{table}
%
\begin{table}
\caption{Summary of hadron masses and bootstrap errors for various
fat-link actions: OFCA ($c_{SW} = 1.1, 1.2$), TFCA ($c_{SW} = 1.0$),
and FWA ($c_{SW} = 0.0$).  \label{summary.tbl}}
\begin{tabular}{ddddddd}
$c_{SW}$ & $6/g^2$ & $\kappa$ & $m_\pi a$ & $m_\rho a$ & $m_N a$ & $m_\Delta a$\\
 \hline
1.1 & 6.0 & 0.1225 & 0.3985(21) & 0.539(5) & 0.807(7) & 0.899(11) \\
1.1 & 6.0 & 0.1230 & 0.3549(23) & 0.511(6) & 0.752(8) & 0.865(14) \\
1.1 & 6.0 & 0.1235 & 0.3072(25) & 0.483(7) & 0.691(9) & 0.832(17) \\
 \hline
1.2 & 5.7 & 0.1200 & 0.731(4) & 0.968(7){\ }  & 1.411(17)  & 1.58(2) \\
1.2 & 5.7 & 0.1220 & 0.569(5) & 0.882(10)     & 1.223(24)  & 1.44(3) \\
 \hline
1.0 & 6.0 & 0.1225 & 0.4293(20) & 0.553(5) & 0.841(7) & 0.920(10) \\
1.0 & 6.0 & 0.1230 & 0.3891(22) & 0.526(5) & 0.791(7) & 0.882(12) \\
1.0 & 6.0 & 0.1235 & 0.3460(25) & 0.498(6) & 0.737(8) & 0.848(14) \\
 \hline
1.0 & 5.7 & 0.1200 & 0.816(4) & 0.992(5){\ } & 1.496(15) & 1.630(19)  \\
1.0 & 5.7 & 0.1220 & 0.682(4) & 0.907(6){\ } & 1.335(16) & 1.503(23)  \\
1.0 & 5.7 & 0.1245 & 0.478(6) & 0.802(10)    & 1.095(28) & 1.341(29)  \\
 \hline
0.0 & 6.0 & 0.1270 & 0.355(3) & 0.455(5) & 0.725(6){\ } & 0.802(13) \\
0.0 & 6.0 & 0.1280 & 0.280(4) & 0.410(5) & 0.631(7){\ } & 0.743(13) \\
0.0 & 6.0 & 0.1290 & 0.186(5) & 0.369(9) & 0.519(13)    & 0.678(22) \\
 \hline
0.0 & 5.7 & 0.1280 & 0.666(5) & 0.780(6) & 1.263(18) & 1.359(17) \\
0.0 & 5.7 & 0.1310 & 0.501(7) & 0.681(8) & 1.073(22) & 1.204(23) \\
0.0 & 5.7 & 0.1330 & 0.435(9) & 0.650(8) & 0.993(22) & 1.151(27) \\
 \hline
\end{tabular}
\end{table}
\begin{table}
\caption{Scaling summary at $m_\pi^2 = 2.5 \sigma$
\label{scalingpi.tbl}}
\begin{tabular}{llllll} \hline
$6/g^2$ & $m_\pi / m_\rho$ & $m_N / m_\rho$ & $m_\rho / \sqrt{\sigma}$
& $m_N / \sqrt{\sigma}$ & $m_\Delta / \sqrt{\sigma}$\\
 \hline
\multicolumn{6}{c}{}\\
\multicolumn{6}{c}{WA}\\
5.7 & 0.815(5) & 1.596(11) & 1.929(21) & 3.075(36) & \ \ \ \ \ $-$\\
6.0 & 0.763(2) & 1.538(6) & 2.072(10) & 3.186(17) & \ \ \ \ \ $-$\\
 & $-$6.4(0.6)\% & $-$3.6(0.8)\% & +7.4(1.3)\% & +3.6(1.3)\% & \ \ \ \ \ $-$ \\
\multicolumn{6}{c}{}\\
\multicolumn{6}{c}{FWA}\\
5.7 & 0.823(11) & 1.604(30) & 1.921(26) & 3.082(59) & 3.362(60) \\
6.0 & 0.768(11) & 1.587(21) & 2.055(23) & 3.261(31) & 3.631(62) \\
 & $-$6.6(1.7)\% & $-$1.1(2.3)\% & +7.0(1.9)\% & +5.8(2.3)\% & +8.0(2.7)\% \\
\multicolumn{6}{c}{}\\
\multicolumn{6}{c}{TFCA}\\
5.7 & 0.704(8) & 1.439(24) & 2.246(29)&  3.233(61) & 3.733(75) \\
6.0 & 0.695(10) & 1.480(23) & 2.275(29) &  3.366(38) & 3.872(65) \\
 & $-$1.3(1.8)\% & +2.8(2.4)\% & +1.3(1.8)\% & +4.1(2.3)\% & +3.7(2.7)\% \\
\multicolumn{6}{c}{}\\
\multicolumn{6}{c}{OFCA}\\
5.7 & 0.678(9) & 1.408(29) & 2.334(34) & 3.285(66) & 3.814(80) \\
6.0 & 0.684(10) & 1.462(24) & 2.310(30) & 3.384(39) & 3.922(68) \\
 & +0.9(2.0)\% & +3.8(2.7)\% & $-$1.0(1.9)\% & +3.0(2.4)\% & +2.8(2.8)\% \\
\multicolumn{6}{c}{}\\
\multicolumn{6}{c}{NPCA}\\
5.7 &  0.671(8)    & 1.432(20)   & 2.356(30)   & 3.375(52)   &  3.854(87) \\
6.0 &  0.674(2)    & 1.488(9)    & 2.345(12)   & 3.490(23)   &     $-$    \\
    &  +0.4(1.2)\% & +3.9(1.6)\% & $-$0.5(1.4)\% & +3.4(1.7)\% &     $-$    \\
 \hline
\end{tabular}
\end{table}

\begin{table}
\caption{Scaling summary at $m_\pi / m_\rho = 0.7$
\label{scalingpirho.tbl}}
\begin{tabular}{lllll} \hline
$6/g^2$ & $m_N / m_\rho$ & $m_\rho / \sqrt{\sigma}$
& $m_N / \sqrt{\sigma}$ & $m_\Delta / \sqrt{\sigma}$\\
 \hline
\multicolumn{5}{c}{}\\
\multicolumn{5}{c}{WA}\\
5.7 & 1.554(16) & 1.698(19) & 2.638(35) & \ \ \ \ \ $-$\\
6.0 & 1.514(8)  & 1.924(10) & 2.913(17) & \ \ \ \ \ $-$\\
 & $-$2.6(1.1)\% & +13.3(1.4)\% & +10.4(1.6)\% & \ \ \ \ \ $-$\\
\multicolumn{5}{c}{}\\
\multicolumn{5}{c}{FWA}\\
5.7 & 1.547(38) & 1.710(27) & 2.646(62) & 3.024(71) \\
6.0 & 1.548(24) & 1.905(24) & 2.952(33) & 3.463(61) \\
 & +0.1(2.9)\% & +11.4(2.3)\% & +11.6(2.9)\% & +14.5(3.4)\% \\
\multicolumn{5}{c}{}\\
\multicolumn{5}{c}{TFCA}\\
5.7 & 1.436(24) & 2.247(29) & 3.235(61) & 3.735(75) \\
6.0 & 1.475(22) & 2.289(28) & 3.393(37) & 3.889(64) \\
 & +2.7(2.3)\% & +1.9(1.8)\% & +4.9(2.3)\% & +4.1(2.7)\% \\
\multicolumn{5}{c}{}\\
\multicolumn{5}{c}{OFCA}\\
5.7 & 1.422(26) & 2.383(33) & 3.393(64) & 3.893(76) \\
6.0 & 1.475(22) & 2.349(29) & 3.465(38) & 3.968(65) \\
 & +3.7(2.4)\% & $-$1.4(1.8)\% & +2.1(2.2)\% & +1.9(2.6)\% \\
\multicolumn{5}{c}{}\\
\multicolumn{5}{c}{NPCA}\\
5.7 & 1.455(9)  & 2.427(10) & 3.532(17) & \ \ \ \ \ $-$\\
6.0 & 1.466(18) & 2.380(17) & 3.488(34) & \ \ \ \ \ $-$\\
 & +0.8(1.4)\% & $-$1.9(1.3)\% & $-$1.2(1.1)\% & \ \ \ \ \ $-$\\
 \hline
\end{tabular}
\end{table}

\begin{table}
\caption{Mass splittings at $m_\pi ^2 = 2.5 \sigma$
\label{splittings.tbl}}
\begin{tabular}{ddd} \hline
$6/g^2$ & $(m_{\rho}^2-m_{\pi}^2)/\sigma$ & $(m_N-m_\Delta) / \sqrt{\sigma}$ \\
 \hline
\multicolumn{3}{c}{}\\
\multicolumn{3}{c}{WA}\\
5.7 & 1.22(3) & $-$\\
6.0 & 1.79(2) & $-$\\
\multicolumn{3}{c}{}\\
\multicolumn{3}{c}{FWA}\\
5.7 & 1.19(7) & 0.28(8) \\
6.0 & 1.72(6) & 0.37(7) \\
\multicolumn{3}{c}{}\\
\multicolumn{3}{c}{TFCA}\\
5.7 & 2.54(7) & 0.50(10) \\
6.0 & 2.68(6) & 0.51(8) \\
\multicolumn{3}{c}{}\\
\multicolumn{3}{c}{OFCA}\\
5.7 & 2.94(8) & 0.53(10) \\
6.0 & 2.84(6) & 0.54(8)  \\
\multicolumn{3}{c}{}\\
\multicolumn{3}{c}{NPCA}\\
5.7 & 3.05(6) & 0.48(5) \\
6.0 & 3.00(2) &  $-$    \\
 \hline
\end{tabular}
\end{table}
\figure{
 \vspace*{-4cm}
 \epsfig{bbllx=200,bblly=130,bburx=830,bbury=940,clip=,
         file=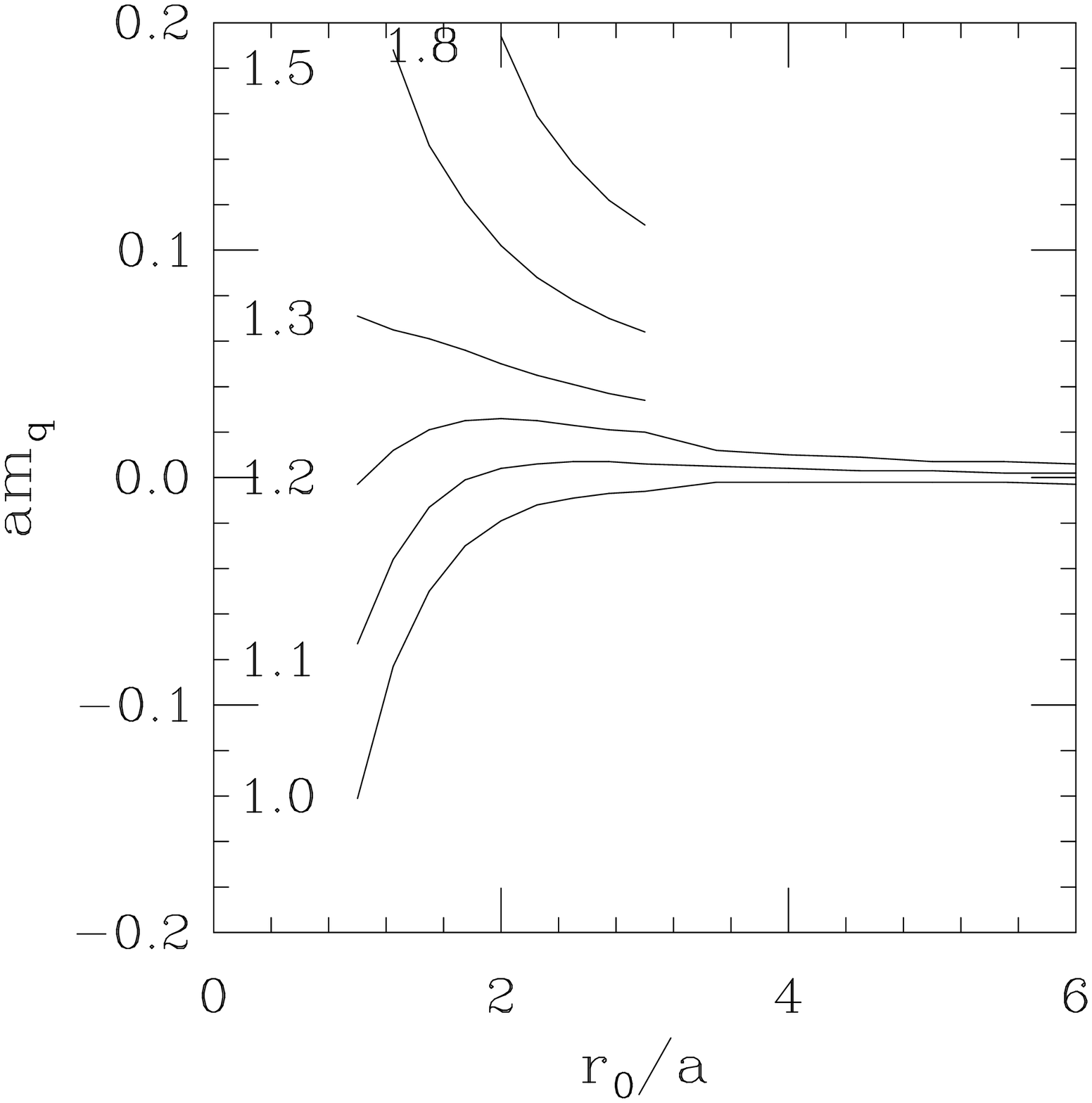,width=160mm}
\caption{Zero mode pole position expressed as a bare quark mass
{\it vs} instanton size for artificial lattice instantons.
Note that all the curves eventually approach
negative infinity for small instanton sizes.
\label{fig:zero_vs_inst_size}
}
}
\figure{
 \vspace*{-4cm}
 \epsfig{bbllx=200,bblly=130,bburx=830,bbury=940,clip=,
         file=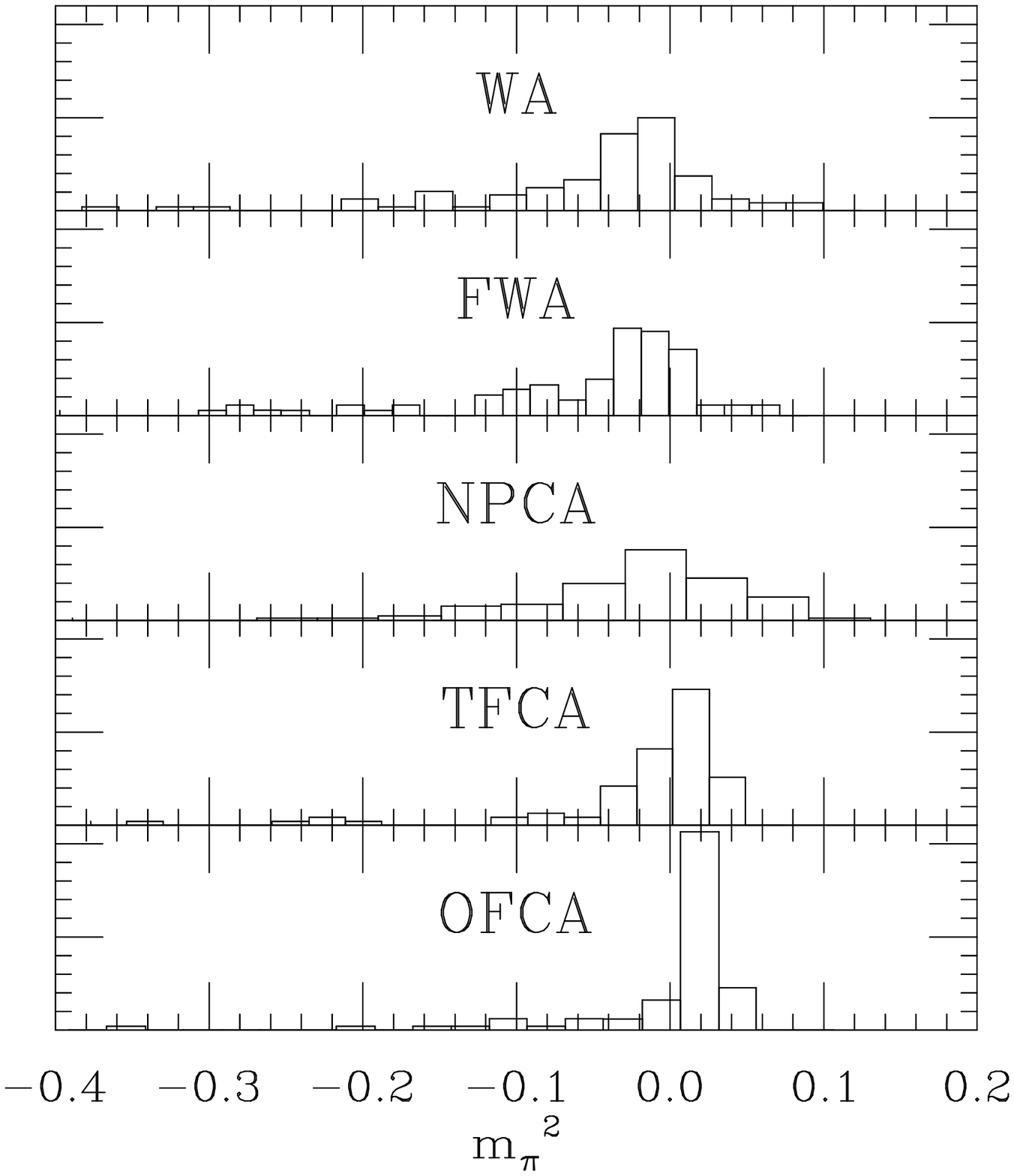,width=160mm}
\caption{Probability distribution of leading eigenvalue for various
fermion actions on 100 $12^4$ gauge configurations at quenched $6/g^2
= 5.85$.
\label{fig:prob_leading}
}
}
\figure{
 \vspace*{-4cm}
 \epsfig{bbllx=200,bblly=130,bburx=830,bbury=940,clip=,
         file=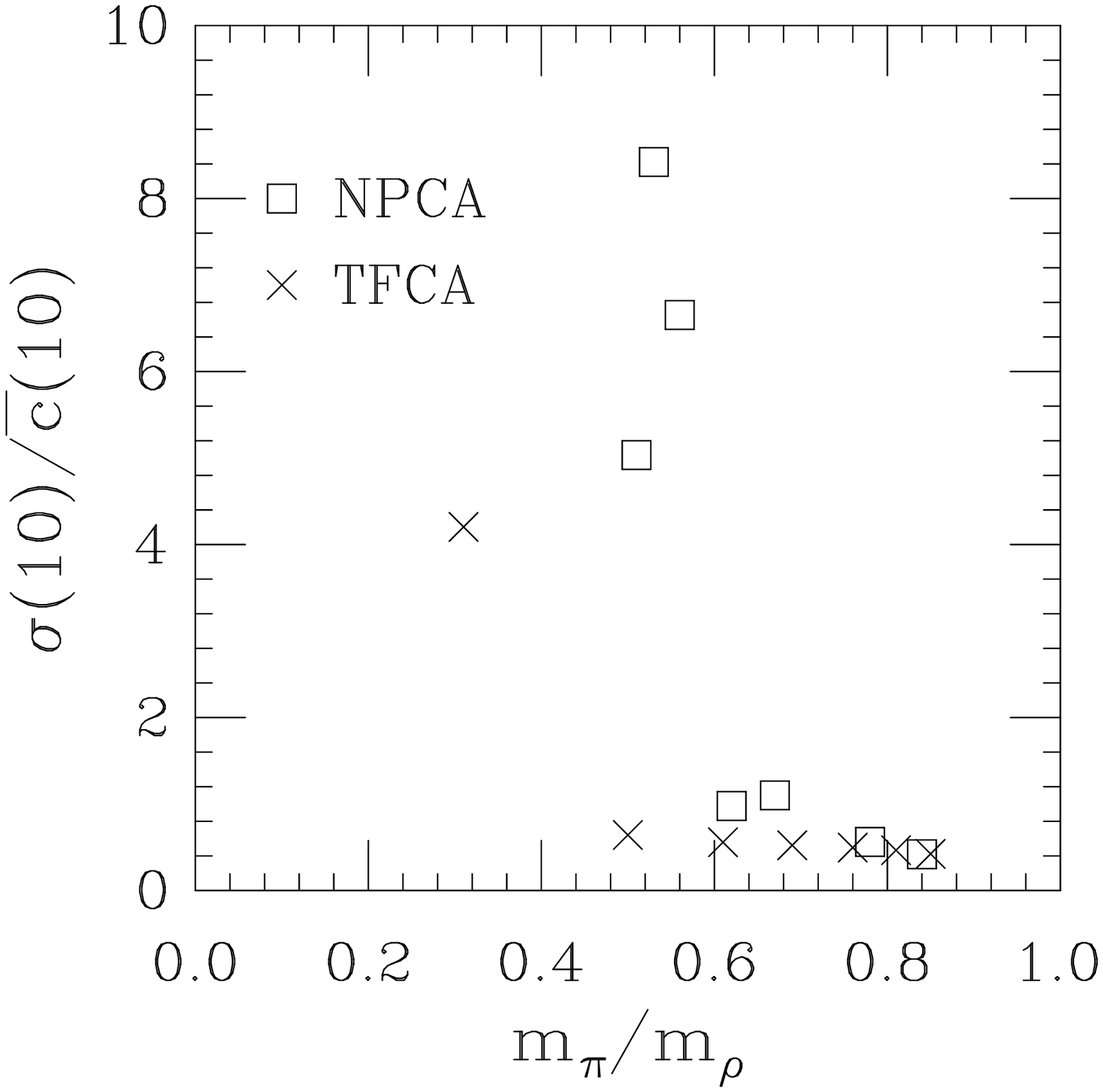,width=160mm}
\caption{Noise to signal ratio for the pion correlator at time $t =
10$ {\it vs} $m_\pi/m_\rho$ for the NPCA and TFCA, on a set of 80
 $8^3\times 24$ lattices at $\beta=5.7$.
\label{fig:noise_to_signal}
}
}
\figure{
 \vspace*{-4cm}
 \epsfig{bbllx=200,bblly=130,bburx=830,bbury=940,clip=,
         file=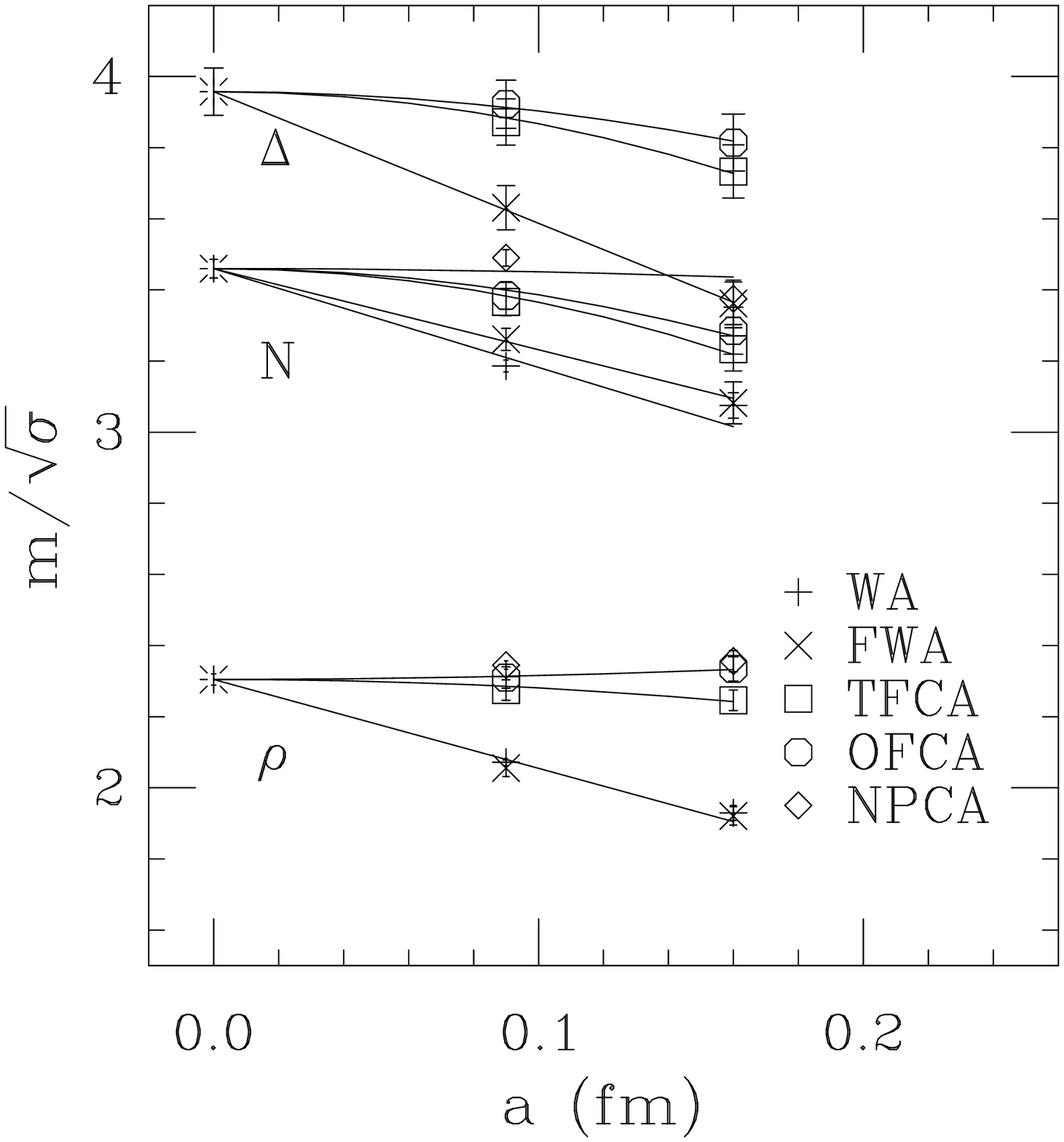,width=160mm}
\caption{Hadron masses in units of the string tension 
for various actions {\it vs} lattice spacing at 
fixed $m_\pi^2 = 2.5 \sigma$.  Masses are extrapolated to a
common continuum value using a function linear in $a$ for WA and FWA
and a function linear in $a^2$ for the other actions.
\label{fig:scalingpimasses}
}
}
\figure{
 \vspace*{-4cm}
 \epsfig{bbllx=200,bblly=130,bburx=830,bbury=940,clip=,
         file=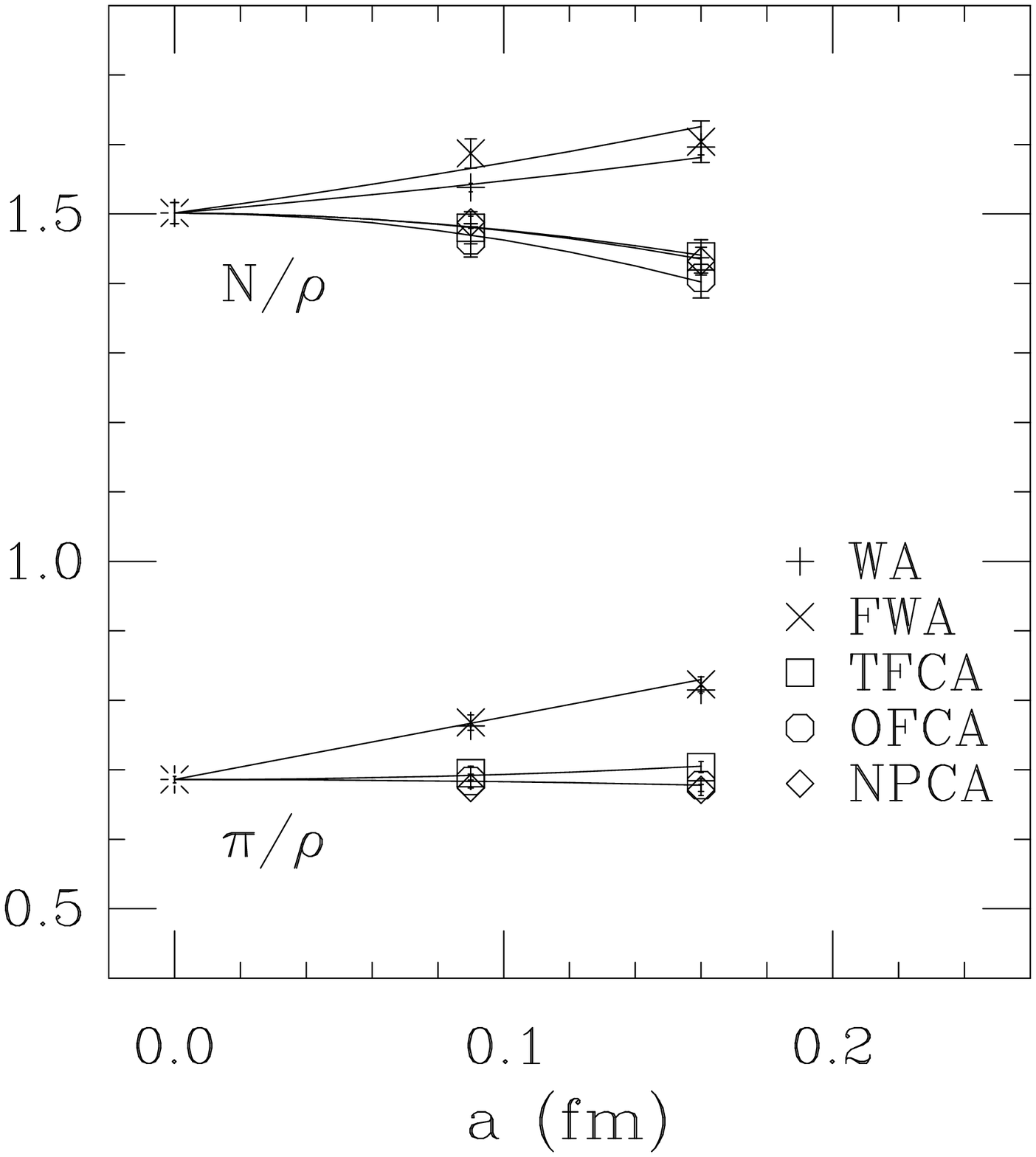,width=160mm}
\caption{Hadron mass ratios for various actions {\it vs} 
lattice spacing at fixed $m_\pi^2 = 2.5 \sigma$.  The extrapolation
is the same as in Fig.~\protect\ref{fig:scalingpimasses}.
\label{fig:scalingpiratios}
}
}
\end{document}